\documentclass[12pt,showpacs]{revtex4}
\usepackage{amsmath}
\usepackage[dvips]{epsfig}

\begin{document}

\title{A Twist (with Pike) for the Simple Harmonic Oscillator.}

\author{G  Filewood}
\email{grf@physics.unimelb.edu.au}
\affiliation{Research Center for High Energy Physics, University 
of Melbourne, Australia.}

\date{\today}
\begin{abstract}
Extension of the formalism of Q.M. to 
resolve mathematical anomalies in 
the structure of anti-unitary 
operators; 
implications for vacuum structure and
spin-statistics arising from an analysis applied
 to the S.H.O. Outline of derived properties
of the S.M. Higgs boson.
\end{abstract}


\maketitle


There exists  an
 anomaly  in the mathematical
 formulation of discrete time reversal  represented
by an anti-unitary operator $\Theta$ 
(and by corollary to anti-unitary operators in 
general);
\[
|\alpha{\cal{i}}\;\rightarrow\;
\Theta|\alpha{\cal{i}}=|\tilde{\alpha}{\cal{i}}
\] 
where $|\tilde{\alpha}{\cal{i}}$ is the time reversed 
state of $|\alpha{\cal{i}}$
(or motion reversed as Sakurai \cite{Sakurai} prefers; whose 
 exposition  we follow).  Comparing the time 
evolved time reversed state in comparison with a ket at an
earlier time;
\[
\left(
1-{\frac{iH}{\hbar}}\delta{t}
\right)
\Theta |\alpha{\cal{i}}
\;=\;
\Theta
\left(
1-{\frac{iH}{\hbar}}(-\delta{t})
\right)
|\alpha{\cal{i}}
\]
from which we deduce;
\[
-iH\Theta|\;{\cal{i}}
=
\Theta
iH|\;{\cal{i}}
\]
which is a problem because it implies;
\[
H\Theta |n{\;\cal{i}}
=
-\Theta
H|n{\cal{i}}
=(-E_n)\Theta{\cal{j}}n{\;\cal{i}}
\]
and the state $\Theta|n{\cal{i}}$ is an eigenket of the Hamiltonian
with negative energy! But the cure is worse than the disease;
\[
\Theta i H|\;{\cal{i}} = -iH\Theta  |\;{\cal{i}}
\;\Rightarrow\;
\Theta H = H \Theta
\]
i.e. assuming the  operator anti-commutes with scalar $i$ 
which is just a number. This cures the problem of negative energy 
but sacrifices mathematical consistency. Indeed,
anti-unitary operators  defined by 
$\Theta^{\dagger}\Theta=-1$ (and of course
$i\Theta|\;{\cal{i}}=-\Theta{i}|\;{\cal{i}}$) cause;
\[
{\cal{h}}\tilde{\alpha}|\tilde{\alpha}{\cal{i}}=
{\cal{h}}\alpha|\Theta^{\dagger}\Theta|\alpha{\cal{i}}
=-{\cal{h}}\alpha|\alpha{\cal{i}}
\]
and in order to avoid states of negative norm
we must invoke an ad-hoc rule that the 
the anti-unitary operator cannot couple to
a bra. 

The cause of these problems is 
the inadmissible of negative norm
or negative energy states into the theory. Indeed, 
Wigner's proof (detailed 
and extended by  Weinberg \cite{Weinberg})
 that symmetry transformations
on a Hilbert space are either unitary and
linear or anti-unitary and anti-linear
requires  
normalized kets ${\cal{h}}\alpha_i|\alpha_j{\cal{i}}=\delta_{i,j}$
as  input because a Hilbert
space with  a positive definite
metric is required for  the probabilistic
interpretation of Q.M. The proof breaks
down if states of non-positive definite norm are
admitted.  An alternative 
approach, however, is to require commuting of scalar numbers
with operators represented as matrices of numbers
and deal with the states of anomalous 
norm when they arise in physical problems
by assuring that they are not observable
so that the probabilistic interpretation
can be  preserved. This paper presents such an approach
using the simple harmonic oscillator as an
example. We shall find this analysis extends our
interpretation of the spin statistics theorem
and has significant implications for our understanding
of the process of spontaneous symmetry breaking
of degenerate vacua leading to the prediction 
of the non-detectability of the Higgs boson.

We begin by 
presenting a 
physical picture  and 
then generating the mathematical structure
that corresponds to the physical picture.
We know that for mass renormalisation 
for a charged object (such as an electron) to 
make sense we must ascribe an infinite negative 
self-mass to cancel the divergent loop
self-energy diagrams. If we believe in
our mathematics then we  want to develop a
theory that hides  states of negative energy
in an `internal' space of such a fermion in such
a way that the negative energy 
states (in analogy to 
Dirac's hole theory) are all filled
but physically unobservable. Now an object
such as an electron is point-like in deep inelastic 
scattering experiments but there is still always 
a definable `internal' momentum space (the object
cannot be squeezed to a geometric point
of zero dimension which would require infinite
energy interaction).
Since the electron is E.M. charged 
 this `internal' momentum space 
is necessarily always supraluminal   i.e. the
interaction  should be point-like
up to any positive energy and the interior space is an
x and p space
of tachyons or tachyonic-like states
where some assumed positive energy 
probe (for example a scattering  photon) 
$E_0^2>0$ in real space-time
defines the (observer dependent) space to which the field is
`confined' $x \approx {\hbar}c{(E_0)^{-1}}$.
The second clue we want to work around 
concerns the Higgs field.
 This starts
life as a tachyon;
\[
{\cal{L}}_{H}=D_{\alpha}\Phi_{(x)}^{\dagger}
D^{\alpha}\Phi_{(x)}
-{\mu^2}|\Phi|^2-{\lambda}|\Phi|^4+g_f(\bar{f}_L
\phi
f_{R}+H.C.)
\]
with $\mu^2<1$
but  with S.S.B. the sign on $\mu^2$ changes.
This is usually interpreted as the conversion 
of the tachyon into a real particle with positive mass
but  the Higgs mass-term sign change could
instead be interpreted as  a tachyon
which  becomes a `hole' of positive energy
in  supraluminal negative 
energy momentum space.
To avoid observation and violation
of the constraints of special relativity
we must then compactify the space in which
the Higgs lives; which we might try to
do by putting it in the un-observable `internal'
momentum  space
of the fermions described above.

The mathematical structure we will use will
consist of a Hilbert space $\Phi$  of real normed bra-kets 
${\cal{h}}\alpha_i|\alpha_j{\cal{i}}=\delta_{i,j}$ 
for the observable states and a 
biquaternionic \cite{Octonions} mirror Hilbert space $\Phi_M$
 of 
pure-imaginary  bra-kets  
for the `internalized' tachyonic states. 
We define a 
biquaternionic mirror Hilbert space $\Phi_M$ as 
a  double-basis  related by mirror
complex conjugation (defined below;
more correctly perhaps we should regard the space of observables
as a rigged-Hilbert space $\Phi_+$ a subset
of   Hilbert space $H$ \cite{Bohm}
and $\Phi_M\subset\Phi_A$
where $\Phi_A$ encompasses a wider class 
of topologies of infinite dimensional
vector spaces including those with 
anomalous norm and $H\subset\Phi_A$).

A conventional Hilbert space 
base ket $|\alpha{\cal{i}}$ 
consists only of a ray \{0,0,0,...0,1,0,...\}.
A mirror  basis ray is composed of the
pair  \{0,0,0,....0,$\frac{I}{\sqrt{2}}$,0,...\} and
\{0,0,0,...0,$\frac{J}{\sqrt{2}}$,0,...\} where I and J are
defined such that $I^2=J^2=K^2=-1$ and
$IJK=-1$ .
An explicit representation is given by the Pauli matrices;
$I=i\tau_1,\;J=i\tau_2,\;\text{and}\;K=i\tau_3$ with the
commutation relation;
$[I,J]_{-}=2\varepsilon_{12}^{\;\;\;\;3}K;\;
\varepsilon_{123}=+1=-\varepsilon_{12}^{\;\;\;3}$.
A base ket
 in Mirror Hilbert Space always has two
rays corresponding to it.
The combination of real and pure imaginary basis
(where each pure imaginary base ket has two components) 
defines a three-dimensional
space of bases; the real basis corresponding to conventional
space-time fields and the dual pure-imaginary basis corresponding
to the `internal' momentum space tachyon fields as we shall see.

 We now
define the operation of mirror conjugation on a matrix M with elements
$a_{i,j}$
which has r rows and c columns
as follows;
\[
M^{\tilde{m}}\;\Rightarrow\;
(a_{i,j})^{\tilde{m}}
\equiv
a^{\tilde{m}}_{(c-j+1),(r-i+1)}
\]
For c-numbers ${\tilde{m}}=*$ where $*$ is the complex
conjugate.
As will be subsequently seen, mirror conjugation is
closely related to the cross product when applied to
the mirror basis kets 
and ${\tilde{m}}$ enables  a generalization
of this  product to be  applied to the mirror basis operators
so in general we restrict its' application to that basis. The 
easiest way to visualize ${\tilde{m}}$ for a square matrix is 
a reflection over the diagonal orthogonal to that upon which
the trace of the matrix is based
followed by mirror conjugate of each element.
 For non-square matrices the
transformation follows the given expression but is
quite simple since, although
each basis is infinite dimensional,  
 we work with two or three component vectors only
under mirror conjugation here. It is 
easy to show that for the 2x2 or 3x3  matrices we work with;
\[
(A+B)^{\tilde{m}}=A^{\tilde{m}}+B^{\tilde{m}}, 
\; (AB)^{\tilde{m}}=B^{\tilde{m}}A^{\tilde{m}}
\]
 In additon to a property like the transpose,
for certain unitary operators $MM^{\dagger}=1$ that are expressible
in terms of quaternions. 
\[
MM^{\tilde{m}}=-1
\]
defines an anti-unitary operator in mirror space. Mirror 
conjugation is one possible solution to the
problems of defining a transpose  \cite{DeLeo} for 
quaternion systems. It is partly motivated by the 
asymmetry between the time and space components
of the E.M.   tensor $F_{\mu\nu}$ paralleling
the lack of symmetry between the cross
$B_{ij}=\nabla_i{\times}A_{j}$ and scalar 
(with vector) $E_{i0}=\nabla_{i}{A_0}$
products
(for constant potential) but we will not
pursue this matter.

Note that with the modified Pauli matrix representation of
the quaternions we have $I^m=-I,\;J^m=J,\;\text{and}
\;K^m=K$. The norm of the mirror bra-kets is;
${\cal{h}}\tilde{\alpha}_i|\tilde{\alpha}_j{\cal{i}}=K$ (in the case of 
discrete eigenvalues; which is the only one we consider
here). Here the forming of a bra-ket may
be taken as the analogue of reducing the
quaternions to the complex numbers and $K$
treated as a pure imaginary number $K \approx i=\sqrt{-1}$
(following Adler \cite{Adler} we might consequently 
expect the associated S-matrix formulation to be complex).
Thus the norm of the mirror kets is pure imaginary and
they are associated with negative probabilities
\footnote{We take a  negative probability  to be a function of a
literal  reverse time process in which cause `follows' effect
and is related to the probability  
that any given initial state $|\tilde{\alpha}_i{\cal{i}}$
out of a set  of  possible such states (the `effects')
leads to  a certain unique  state  $|\tilde{\alpha}_f{\cal{i}}$
(the `cause'),  
with time flowing backwards and $P= 
{\cal{h}}\tilde{\alpha_f}|{\cal{T}}|\tilde{\alpha}_i{\cal{i}}
{\cal{h}}\tilde{\alpha_i}{\cal{i}}|
{\cal{T}}^m|\tilde{\alpha}_f{\cal{i}}
\leq0$ where ${\cal{T}}$ is the appropriate reverse time evolution operator.}.

An operator $A$ is defined as `mirror-hermitian' 
(res. mirror-anti-hermitian) in the supraluminal
space of mirror kets
if $A^m=-A$ (res. $A^m=+A$) and a mirror ket is an eigenket of the
operator if $A|\tilde{\alpha}{\cal{i}}=a'|\tilde{\alpha}{\cal{i}}$
for some quaternion number $a'$ and
$\int|\tilde{\alpha}{\cal{i}}{\cal{h}}\tilde{\alpha}|=e^{i\theta}K$
defines a complete set of states for $\theta$ a real parameter
dependent upon the way we define the mirror basis pair.

The connection between mirror and real Hilbert spaces
is given by the time reversal operator $\Theta$;
and its action on a purely real-basis ket $|\alpha{\cal{i}}$ is to produce
the analogous pure imaginary base ket $|\tilde{\alpha}{\cal{i}}$.
This involves the generation of a two-component base ket from
a unicomponent base ket and the entire Hilbert space must be 
split into two pieces. This may  require  modeling
in terms of fields with deconvolution integrals but a simple
approach is to
take $\Theta|\;\;{\cal{i}}$ as
\footnote{$\Theta$ must be constructed using the same
negative energy basis  as the 
$|\tilde{\alpha}{\cal{i}}$; thus if we 
interchange the $I$'s and $J$'s or their respective 
signs / phase  in the definition of the 
basis kets then we do the same for the time reversal operator.};
\begin{eqnarray}
\left(
\begin{array}{ccc}
0&\frac{I}{\sqrt{2}}&0 \\
\frac{-I}{\sqrt{2}}&0&\frac{-J}{\sqrt{2}} \\
0&\frac{J}{\sqrt{2}}&0
\end{array}
\right)
\left(
\begin{array}{c}
|\frac{I}{\sqrt{2}}n{\cal{i}}\\
|\;\;\;n\;\;{\cal{i}}\\
|\frac{J}{\sqrt{2}}n{\cal{i}}
\end{array}
\right)
\end{eqnarray}
And it is easy to show that ${\cal{h}}\alpha|\Theta^m
\Theta|\alpha{\cal{i}}={\cal{h}}\tilde\alpha|
\tilde\alpha{\cal{i}}$ and
${\cal{h}}\tilde{\alpha}|\Theta^m
\Theta|\tilde{\alpha}{\cal{i}}={\cal{h}}\alpha|
\alpha{\cal{i}}$
although $\Theta$ is not formally
an invertible  matrix.
(Note that $|\;\;{\cal{i}}^m={\cal{h}}\;\;|
={\cal{h}}\frac{J}{\sqrt{2}}n|{\cal{h}}n|{\cal{h}}
\frac{-I}{\sqrt{2}}n|$).

Now consider the following 
analogues of the S.H.O. creation and annihilation
operators;
\[
a\Rightarrow A=
\left(
\begin{array}{ccc}
0&0&{ia}\\0&{a}
&0\\{-ia}&0&0
\end{array}
\right)\;;
\;A^{m}=
\left(
\begin{array}{ccc}
0&0&{-ia^{m}}
\\0&a^{m}&0\\{ia^m}&0&0
\end{array}
\right)
\]
and imposing the  non-trivial constraint
$a^m=a^{\dagger}$\footnote
{Because the $A$ and $A^m$ do not mix the real and imaginary states
the two sets of operators could in principle 
be separated so that $a^m$ and $a^{\dagger}$ need not be equal;
however it is not clear in that circumstance that the vacuum
energy would vanish.}
now let us apply these operators respectively to
energy eigenkets (negative for the imaginary states);
\begin{eqnarray}
A\;|n{\cal{i}}&=&{\sqrt{n}}\;|n-1{\cal{i}}\nonumber
\\ \nonumber
A\;|\frac{I}{\sqrt{2}}{{(-n)}}{\cal{i}}
&=&{\frac{-iI}{\sqrt{2}}}{\sqrt{(-n)}}\;|{\frac{I}{\sqrt{2}}}
[-(n-1)]{\cal{i}}
\end{eqnarray}
where we have used the definition of the mirror kets as
a `normalization'
and similarly for the $J$ mirror 
component. Thence subsequently applying $A^m$ we obtain;
\begin{eqnarray}
A^m{\sqrt{n}|n-1{\cal{i}}={{n} }}\;|n{\cal{i}}
\nonumber \\
A^m\frac{-iI}{\sqrt{2}}{\sqrt{(-n)}}\;
|\frac{I}{\sqrt{2}}[-(n-1)]{\cal{i}}
&=-\frac{n}{2}\;|\frac{I}{\sqrt{2}}(-n){\cal{i}}
\end{eqnarray}
and vis-a-vis for the $A^mA|\frac{J}{\sqrt{2}}(-n){\cal{i}}
=-\frac{n}{2}|\frac{J}{\sqrt{2}}(-n){\cal{i}}$ so the
 occupation number (or energy)
of the imaginary state $|\tilde{n}{\cal{i}}$ is
-n. We have (with the constraint that 
$a^{\dagger}=a^m)$;
\begin{equation}
A^{\tilde{m}} A =
\left(
\begin{array}{ccc}
-a^{\dagger}a&0&0\\
0&a^{\dagger}a&0 \\
0&0&-a^{\dagger}a
\end{array}
\right)={\mathcal{H}}
\label{Hamiltonian}
\end{equation}
In standard S.H.O. theory
${\mathcal{H}}= a^{\dagger}a + 1/2$ but 
here we expect the true
 vacuum energy (the vacuum disassociated 
from the oscillator) to be  zero
since the two $-a^{\dagger}a$'s in this Hamiltonian
are associated with the same occupation number as
the single $+a^{\dagger}a$ but with  opposite energy.
Note that;
$
{\mathcal{H}} \Theta 
=
-\Theta {\mathcal{H}}
$
so we have avoided the requirement for anti-linearity.

Since the number operator $a^{\dagger}a$ has changed sign
for the mirror components we have;
$
[x,p]_-=i\hbar
\Rightarrow
[\tilde{x},\tilde{p}]=
-i\hbar
$
and as  a consequence the commutator of the bosonic
creation and annihilation operators in the pure
imaginary basis changes sign also;
\begin{equation}
[a,a^{\dagger}]_-|\alpha{\cal{i}}=1|\alpha{\cal{i}}
\Rightarrow[\tilde{a},\tilde{a}^{\dagger}]_-|\tilde{\alpha}{\cal{i}}
=
-1|\tilde{\alpha}{\cal{i}}
\label{comm}
\end{equation}
when applied to pure imaginary kets $|\tilde{\alpha}{\cal{i}}$
(as can be proved directly from the definition of the kets also).
It is easy to show that the states of negative energy
(or negative occupation number; the physical meaning of which
we will investigate later) obey Bose-Einstein statistics
as expected from the commutation relation 
$[\tilde{a},\tilde{a}^{\dagger}]_{-}=-1$. However,
we are interested in a positive energy `hole' in
supraluminal space. The theory admits such a `hole';
but only as a complex doublet; holding the definition 
of $\Theta$ fixed in the negative energy basis we have
the definition of positive energy `holes' in the
negative basis as;
\begin{equation}
\left(
\begin{array}{c}
|\frac{+iI}{\sqrt{2}},-1{\cal{i}}\\
|\;\;\;0\;\;{\cal{i}}_R\\
|\frac{-iJ}{\sqrt{2}},-1{\cal{i}}
\end{array}
\right)={\phi^{+}}
\;\;\text{and}
\;\;
\left(
\begin{array}{c}
|\frac{-iI}{\sqrt{2}},-1{\cal{i}}\\
|\;\;\;0\;\;{\cal{i}}_R\\
|\frac{+iJ}{\sqrt{2}},-1{\cal{i}}
\end{array}
\right)={\phi^{-}}
\label{CSF}
\end{equation}
both  with positive occupation number / energy under
$\tilde{N}=-\tilde{a}^{\dagger}\tilde{a}$. 
(Phase changes to the imaginary basis kets are of
the form $e^{i\theta\cdot\tau}=e^{\theta\cdot{N}}$
where $N=\{I,J,K\}$ and here we take the phase
$e^{\theta\cdot{K}}|\tilde{\alpha}{\cal{i}}=
|e^{i\theta}\frac{I}{\sqrt{2}},n{\cal{i}},
|e^{-i\theta}\frac{J}{\sqrt{2}},n{\cal{i}}$
with diagonal $K$ and $\theta=\frac{\pi}{2}$).

 However the spin statistics
 of the positive energy `hole' are not the same
$\tilde{N}|\phi^{\pm},-1{\cal{i}}=+1
|\phi^{\pm},-1{\cal{i}}$
whilst (using eq.(\ref{comm})); 
\[
\tilde{N}\tilde{a}^{\dagger}|\phi^{\pm},-1{\cal{i}}=0
\;\;\text{and}\;\;
\tilde{N}\tilde{a}|\phi^{\pm},-1{\cal{i}}=0
\]
and the positive energy scalar field 
`holes' in supraluminal momentum space
obey Fermi-Dirac `ghost' statistics;
this is the reason for the choice of $n=-1$
 in
the above definition. This is  an important 
result so let us  derive it by a second route.

A scalar field obeying Fermi-Dirac
statistics appears to be a  violation of the spin-statistics 
theorem. However,
 for the supraluminal 
oscillator, `causality' 
(or more correctly perhaps `reverse-causality')
involves only states that always
have space-like separation. Thus we require
vanishing of field commutators / anti-commutators
for {\it{time-like}} separations (in contrast
the the requirement for {\it{space-like}}
separations
for conventional fields). Consider the scalar 
field;

$\;\;\;\;\;\;\;\;\;\;\;\;\phi(x)\equiv \kappa \phi^{+}(x)
+
\lambda\phi^{-}(x)$

\noindent
for space-like separations $(\delta{x})^2<0$ conventionally
we have \cite{Weinberg};
\begin{eqnarray}
[
\phi^{}_{(x)},\phi^{\dagger}_{(y)}]_{\mp}&=&
|\kappa|^2[\phi^+_{(x)},\phi^-_{(y)}]_{\mp}
+
|\lambda|^2[\phi^-_{(x)},\phi^+_{(y)}]_{\mp} \nonumber \\ 
&=&(|\kappa|^2{\mp}|\lambda|^2)\Delta_+(x-y)
\label{equal-time}
\end{eqnarray}
and we require the  - commutator sign to enforce causality
(and vis-a-vis for the commutator;
$[
\phi_{(x)},\phi_{(y)}]_{\mp}$) i.e. Bose-Einstein statistics. 
In supraluminal mirror space microcausality (see footnotes)
correspondingly requires vanishing of the commutators
in that part of space-time inaccessible to the 
mirror states (i.e. the sub-luminal momentum
space); and we replace the equal time 
delta function by an equal space delta function
(time-only separation and thus clearly Lorentz invariant);
\[
\Delta_{-}{(t)}=\frac{1}{2\pi}\int{dp_0}\,
\epsilon(p_0)\,e^{-ip_0\cdot{t}}
\]
where $\epsilon(p_0)$ changes sign with time-ordering.
Now eqs.(\ref{equal-time}) are equal-time and not in 
normal ordering; 
\begin{eqnarray}
[
\phi^{}_{(t)},\phi^{\dagger}_{(t')}]_{\mp}&=&
|\kappa|^2[\phi^+_{(t)},\phi^-_{(t')}]_{\mp}
+\mp
|\lambda|^2[\phi^+_{(t')},\phi^-_{(t)}]_{\mp} \nonumber \\ 
&=&(|\kappa|^2{\pm}|\lambda|^2)\Delta_{-}({t-t'})
\end{eqnarray}
and so now we must take the anti-commutators (the
lower combination of  signs) and the 
scalar field in supraluminal space obeys Fermi-Dirac
statistics. This is the same result we obtained 
previously but through a different route.

Notice that there is a degeneracy  in the definition
of the supraluminal imaginary basis depending upon
the mixture of $I$ and $J$ in each 
of the two vector components (phase factors of the 
form $e^{\sigma{I}}$ and $e^{\epsilon{J}}; \sigma,\epsilon$
real, mix the 
components). This implies that the vacuum
is degenerate for the supraluminal component and
the overlapping vacua of the real and pure imaginary basis
will invoke spontaneous symmetry breaking
of the complex scalar field eq.({\ref{CSF}}) which 
clearly has a non-vanishing 
VEV for the real-space vacuum $|0\cal{i}_R$.
In the S.M., 
S.S.B. of a complex scalar field results 
from the choice of potential which induces
degeneracy of the vacuum; here it arises as a consequence 
of the   
requirement for consistency in the
definition of anti-unitary operators.

To interpret this structure we follow closely in the
footsteps of Dirac's `hole' theory. Since the negative 
energy states are not observed we  interpret 
all the negative energy Bose-Einstein de-excitations
as filled
\footnote
{The negative energy states $|\tilde{\alpha}{\cal{i}}$
are associated with
 negative probabilities
but  the states
which are  preferentially occupied are those 
closest to the vacuum state $|\tilde{0}{\cal{i}}$
which is the {\it{highest}} energy rather than 
the lowest because  the flow of probability 
(or entropy) is reversed;
interestingly this also means that `charged' tachyon
 states, if they existed, should not radiate Cherenkov radiation
unless they transgress into the sub-luminal momentum space
of real kets.}; 
 this will contribute a $-\infty$ 
energy to the pure imaginary basis vacuum
of the `internal' momentum space of a massive 
fermion  which we must normalize away
by canceling it with gauge boson ultraviolet
divergences. 

If a negative occupation number means that 
the tower of states is full then ipsi-facto
a positive occupation number in mirror momentum
space means that it is a `hole' in that space
and equivalently the Higgs becomes confined 
to the supraluminal internal momentum space
because it has no definition in the space
of real kets. This leads to the prediction that
the Higgs will not be seen at the L.H.C. as a free
particle.

In the S.M. the gauge boson / gauge boson
 scattering amplitude
$V_LV_L\rightarrow{V_LV_L}$ contains pieces related to
exchange of electro-weak bosons, the  ($V,\gamma$) vertex,
and Higgs exchange ($V,H$) vertex and both grow 
as $\sim{E}^2$
but their $E^2$ contributions cancel and the cross-sections
is dominated by exchange of transversely polarised
bosons and
grows only as $E^0$ preserving S-matrix unitarity \cite{He};
so the virtual Higgs is needed to preserve unitarity.
 The Higgs model presented in this paper 
demands an `inverted' ghost tachyonic 
geometry in the vicinity of the 
$W^{\pm}$ and the $Z^{0}$ bosons (the latter as a function
of $sin^2\theta_w$); and a preliminary analysis suggests
 these must
be identified  with the t'Hooft gauge
 $\xi<<\infty$ 
that is, with a non-unitary gauge. These matters will be
pursued elsewhere. A viable theory
must preserve the unitarity constraints of
the standard model; naively  we might
expect this to be the case here since the 
S-matrix is independent of the choice of  gauge.

In summary, in spite of the
many simplifications employed
in this presentation,  we see yet again that the S.H.O. 
is an extraordinarily rich theoretical
laboratory. Taken at face value, reforming
our approach to anti-unitary operators appears
capable of providing   new insights into the process of S.S.B.
with the standard model Higgs.
Clearly a more complex theory is required to
describe physical structure as formal fields
\cite{Adler}. The chief prediction of the 
theory is  that the Higgs
boson will not be detectable as a free particle
but instead enjoys an `occult' existence.

\end{document}